# A Neuro-Fuzzy Method to Improving Backfiring Conversion Ratios

Justin Wong[1], Danny Ho[2], Luiz Fernando Capretz[1]
*jwong343@uwo.ca, danny@nfa-estimation.com, lcapretz@eng.uwo.ca*
*1. Department of Electrical & Computer Engineering*
*University of Western Ontario*
*London, Ontario, N6A 5B9, Canada*
*2. NFA Estimation Inc.*
*London, Ontario, N6G 3A8, Canada*

## Abstract

*Software project estimation is crucial aspect in delivering software on time and on budget. Software size is an important metric in determining the effort, cost, and productivity. Today, source lines of code and function point are the most used sizing metrics. Backfiring is a well-known technique for converting between function points and source lines of code. However when backfiring is used, there is a high margin of error. This study introduces a method to improve the accuracy of backfiring. Intelligent systems have been used in software prediction models to improve performance over traditional techniques. For this reason, a hybrid Neuro-Fuzzy is used because it takes advantages of the neural network's learning and fuzzy logic's human-like reasoning. This paper describes an improved backfiring technique which uses Neuro-Fuzzy and compares the new method against the default conversion ratios currently used by software practitioners.*

## 1. Introduction and Background

Software estimation is a delicate activity because it can lead to projects' success or failure. Underestimation can result in poor quality software because of pressure to complete the project and premature release of the software. Furthermore, underestimation may result in late delivery or even failure because there may be lack of resources or budget to complete the project within the delivery date. Overestimation may result in failure to win bids for client projects. It can also result in allocating too many resources and underused staff [1]. Many software estimation techniques such as Constructive Cost Model (COCOMO) and Function Point Analysis have been developed to tackle these problems [2].

The two popular sizing metrics used today are source lines of code (SLOC or LOC) and function points. Jones [3] introduced the concept of backfiring, which converts function points to logical SLOC statements to effectively sized source code. While backfiring is useful and simple, there is a high margin of error in converting SLOC data into function points [3]. Conversion ratios have a range of values, i.e., low, mean, and high; hence, it is difficult to figure out the precise conversion ratio to use for a specific organization.

Neural networks and fuzzy logic have been applied to software estimation methods to improve the accuracy of existing software sizing methods [4, 5]. The neuro-fuzzy technique, which integrates neural network and fuzzy logic together, has been shown to improve software estimation techniques by calibrating its parameters. Based on the flexibility of the technique, the neuro-fuzzy framework was incorporated to the backfiring technique.

The objective of this study is to improve the accuracy of backfiring by creating a neuro-fuzzy function point backfiring (NFFPB) calibration model. The model would be used to size software applications when the programming language and function point are known. It would also be used to provide up-to-date and calibrated conversion ratios to improve performance for specific datasets.

### 1.1. SLOC

SLOC is one of the most popular size metrics used today in industry. SLOC measures the lines of code in software. SLOC is commonly used because it is relatively easy to measure and count. Typically SLOC is used to determine development effort required for software.

While SLOC is popular and simple, there are some shortcomings to the use of SLOC as a size metric. SLOC is not a reliable metric during the software development process because the SLOC often changes from source code being added and removed. Moreover, the level of staff programming skill affects the SLOC produced. For example, an experienced programmer may write less SLOC than a junior programmer. Finally, SLOC is a risky metric if used alone because it is difficult to count an application during development. SLOC counts are more effective when development is completed [6].

### 1.2. Function Point

First introduced in the 1970s by Albrecht [7], the function point is a unit of measurement for determining the functional size of an information system. Function Point Analysis (FPA) can be used to calculate the size of an application, project development or enhancement project. Calculation of application means the size of an existing application which the functionality available to the end users are measured. For a development project, the measure of the size of the new application being developed is considered based on the requested requirements. Calculating an enhancement project measures a maintenance project of a present application based on all changes, added and removed functions, and the data conversion process [8].

To avoid the shortcomings of SLOC, function points are used. However, function points still have their disadvantages. For example, functionality in legacy systems are forgotten, therefore the overall size of the system is underestimated. Moreover, incomplete or informal requirements would result in inconsistent function point counts. Using function point in analogies can result in wrong estimates because a 1000 function point system does not mean it is two times larger than a 500 function point system [6].

### 1.3. Backfiring

The backfiring technique is a bidirectional mathematical conversion between function point and SLOC. The equation for converting function point to SLOC is defined in (1). In equation (1), FP is the function point input and conversion is the SLOC/FP. Equation (1) can be rearranged to find the number of function points if the SLOC is known.

$$SLOC = FP \times Conversion \quad (1)$$

For the backfiring technique, the conversion ratios or SLOC/FP could be found in a programming table. A programming table is tabulated data of the programming language, language level, and SLOC/FP. A sample programming table is shown in Table 1. The language level, defined by Software Productivity Research (SPR), indicates how powerful the source code language is [9]. For example, Basic Assembler has a language level of 1.0 because it has an average of 320 SLOC/FP. The SLOC/FP decreases as the language level increases, which is shown in Figure 1.

*Table 1 – Programming Language Table*

| Language | Language Level | SLOC/FP | | |
|---|---|---|---|---|
| | | Low | Mean | High |
| Basic Assembly | 1.0 | 213 | 320 | 427 |
| C | 2.5 | 21 | 128 | 235 |
| Cobol | 3.0 | 65 | 107 | 170 |
| 3$^{rd}$ Generation | 4.0 | 45 | 80 | 125 |
| C++ | 6.0 | 30 | 53 | 125 |
| Java | 9.0 | 20 | 36 | 51 |
| 4$^{th}$ Generation | 16.0 | 16 | 20 | 24 |
| SQL | 25.0 | 8 | 13 | 17 |

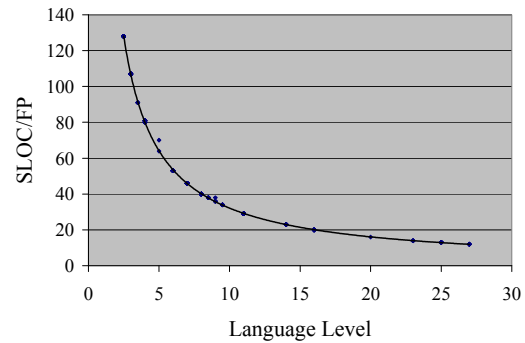

*Figure 1 - Conversion ratios versus language level*

### 1.4. Fuzzy Logic

Fuzzy logic, used to generate a mapping between input and output spaces, is derived from the fuzzy set theory, which uses linguistic terms or a fuzzy set that represents and processes uncertainty. It is made up of membership functions, used to describe linguistic terms such as low, medium and high, with values ranging from 0 to 1.

### 1.5. Neural Network

A neural network is also known as a parallel distribution processing network. Neural network loosely models after the human brain in order to achieve human-like performance and learning. A neural network is made up of a network architecture, activation function, and learning mechanism.

Neural networks have been applied to enhance traditional software estimation techniques. Aggarwal et al. [10] presented a model using neural networks to improve SLOC estimates when the function point is given as an input. Their estimation model used Bayesian Regularization to train the neural network. They investigated on various training algorithm to find the best results. Furthermore, the network took into account the maximum team size, function point standard and language (3rd generation language and 4th generation language). The shortcoming of the neural network was that it had a black-box design. In addition, the research only took the generation language into account instead of the programming languages. The averages SLOC/FP between the 3rd and 4th generation languages are very large in range. The 3rd generation default language has 80 SLOC/FP, while the 4th generation default language has 20 SLOC/FP.

### 1.6. Neuro-Fuzzy Integration

Neuro-Fuzzy is a term used to refer to a hybrid intelligent system using both neural network and fuzzy logic. Neuro-Fuzzy is used because it takes advantages of the neural network's learning and fuzzy logic's human-like reasoning [4].

Huang et al. [4] introduced a neuro-fuzzy framework and applied it to COCOMO. The neuro-fuzzy technique showed it could be used to improve software estimation techniques by calibrating its parameters. In this study, it was demonstrated that a neural network could be used to calibrate fuzzy sets to improve performances for many different applications. Based on the flexibility of the technique, the neuro-fuzzy framework was applied to the backfiring technique in this thesis.

Xia [5] proposed a neuro-fuzzy function point model for improving the accuracy of software effort estimates. The model used fuzzy logic for the function type complexities. In addition, a neural network was used to calibrate the complexity to achieve more accurate and current complexity ranges. The neuro-fuzzy function point model was evaluated and found to have an improvement of 22% over the traditional function point complexity values.

## 2. Neuro-Fuzzy Backfiring Model

The NFFPB model is an approach that calibrates the programming language level to improve backfiring estimations. It uses fuzzy logic to model the programming language level curve by grouping the programming language levels into fuzzy sets. These fuzzy sets become the input programming language level. A neural network is used to calibrate the fuzzy sets' SLOC/FP. The calibrated SLOC/FP replaces the original backfiring conversion ratios.

### 2.1. Fuzzy Programming Language Level

The approach to model the inverse curve was to group the language levels into various fuzzy levels. This was done by grouping all the programming languages together based on similar SLOC/FP. The number of fuzzy level depends on the number of data points available. For example, if there are many points available then more fuzzy levels can be created. Table 2 shows the range of language levels being grouped into a fuzzy level to approximate the curve. The fuzzy levels were obtained based on the programming languages and data points available from ISBSG [11]. The average SLOC/FP was obtained from the average of all the backfiring conversion values within a fuzzy level. Moreover, this average value was used as the initial weight in the neural network and the initial peak of the fuzzy membership functions.

The fuzzy levels would change based on the data available. If more programming languages were to become available, more fuzzy levels could be added to model the curve more accurately.

*Table 2 – Sample of fuzzy levels based on ISBSG Research Suite Release 9*

| Fuzzy Level | Programming Language Level Range | Average SLOC/FP |
|---|---|---|
| 1 | (0..2.5] | 128 |
| 2 | (2.5..3] | 107 |
| 3 | (3..3.5] | 91 |
| 4 | (3.5..4] | 81 |
| 5 | (4..5] | 67 |
| 6 | (5..6] | 53 |
| 7 | (6..7] | 46 |
| 8 | (7..8] | 40 |
| 9 | (8..8.5] | 38 |
| 10 | (8.5..9] | 36 |
| 11 | (9..9.5] | 34 |
| 12 | (9.5..11] | 29 |
| 13 | (11..14] | 23 |
| 14 | (14..16] | 20 |
| 15 | (16..20] | 16 |
| 16 | (20..23] | 14 |
| 17 | (23..25] | 13 |
| 18 | (25..27] | 12 |
| 19 | (27..50] | 6 |

### 2.2. Input and Output Membership Functions

Figure 2 and Figure 3 illustrate the input and output fuzzy membership functions. A triangular function was used for both the input and output membership functions. The peak of the input triangle of each fuzzy level was the programming language level. The average SLOC/FP was the peak of the output membership functions. The peaks were obtained from Table 2.

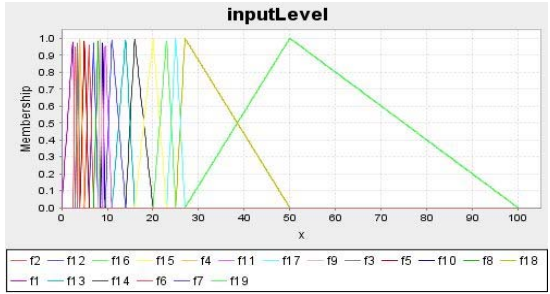

*Figure 2 – Fuzzy membership of the language levels*

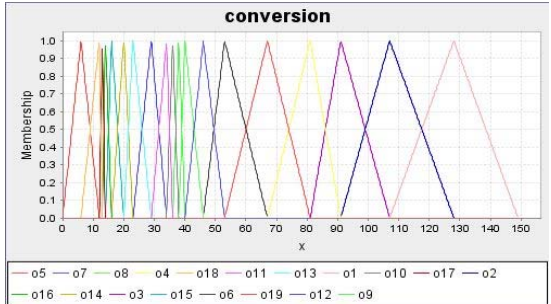

*Figure 3 – Fuzzy membership of the output SLOC/FP*

### 2.3. Fuzzy Rules and Inference

Each fuzzy level was directly referenced to a fuzzy output. For example, f1 referenced o1, f2 referenced o2 and so on. The "AND" and "activation function" used the minimum function for the rules. For defuzzification, the maximum accumulation method and "Center of Gravity" method were used.

### 2.4. Neural Network

The neural network was used to calibrate the average source statements per function point for each fuzzy level. Moreover, the input to the neural network was the UFP and the fuzzy language level. The actual SLOC was used only during training. These fuzzy language levels are fed into the network shown in Figure 4, which shows the network's design. The neural network was designed to be easily interpreted so that it avoids being an obscure "black-box" model. Furthermore, the network design has been shown by Xia [5] to have improvements in function point estimations.

The $L_1$ to $L_n$ were binary fuzzy language level inputs. When a language level was fed into the network, the input was in the form of a matrix and only contains one 1 entry. For example, for language level 4, it would be represented as [0 0 0 1 0 0 0 0 0 0 0 0 0 0 0 0] based on the proposed fuzzy levels.

The weights were initialized initially with the values in Table 2. The step-by-step pseudo-code of how the network is trained is shown below:

1. Input: UFP, Fuzzy Programming Language Level ($L_x$)
2. Input goes through node Z's activation function: *Prediction*= UFP x ($L_x$ x $W_x$)
3. Get error: *Error = Actual - Prediction*
4. Propagate error back and adjust weights: $W_x' = W_x + \eta Error$
5. If $W_x'$>max (SLOC/FP for $L_x$) then $W_x'$= max (SLOC/FP for $L_x$).
   If $W_x'$<min (SLOC/FP for $L_x$) then $W_x'$= min (SLOC/FP for $L_x$).
6. Error goal or epochs reached? If Yes END, else go to step 1.

There is a constraint to how much a conversion weight can change for each fuzzy level. Each fuzzy level's input weight is constrained between the lowest and highest conversion value to avoid the SLOC/FP to be calibrated out of the language's boundaries.

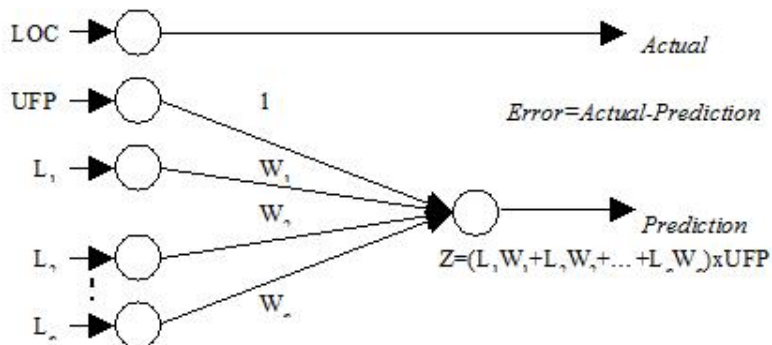

*Figure 4 – Neural Network Design*

## 3. Method of Assesment

The Neuro-Fuzzy Function Points Backfiring (NFFPB) model was implemented into a tool used to calibrate the backfiring conversion ratios. The NFFPB model was trained and evaluated with the International Software Benchmarking Standards Group (ISBSG) Release 9 data repository [11]. The NFFPB performance was measured against the Software Productivity Research (SPR) backfiring conversion ratios.

### 3.1. Evaluation Criteria

Magnitude of Relative Error (MRE) is a popular criteria used commonly in industry. In the evaluation, MRE is used to measure the error performance for the calibrated model against the original conversion ratios. While it is popular, Foss *et al.* [12] showed that when evaluating and comparing prediction models, MRE is not recommended because MRE does not prove that one model is particularly better than another because the results were misleading [13]. MRE favored underestimation and performed worse in small sized projects. The equation is defined in (3).

$$MRE = \frac{|Actual - Predicted|}{Actual} \quad (3)$$

Another method that was proposed for evaluating and comparing prediction models was Magnitude of error Relative to the Estimate (MER) [13]. The equation for calculating MER is defined in (4). However, MER favors overestimation because the estimation was a divisor; therefore, larger estimates tend to perform much better than small estimates.

$$MER = \frac{|Actual - Predicted|}{Predicted} \quad (4)$$

Another criterion that was used to evaluate the prediction model was Prediction at Level (PRED). MRE less than 25% and 50% was utilized for PRED because other models have used these criteria for evaluation.

### 3.2. Experiments

There were 7 different experiments conducted. In each experiment, the MMRE, MMER, and PRED were compared with the original and calibrated ratios.

In experiments 1 and 2, half of the dataset was trained and the rest was used for simulation and evaluation. The data points were selected randomly for each programming language level.

Experiments 3 and 4 were based on the size of the projects. In experiment 3, the large function point count projects were used for training and the small function point counts were used for simulation. Similarly, experiment 4 used small projects for training and large projects for evaluation.

Experiments 5, 6, and 7 were based on a larger training set to see if the performance improves. Experiment 5 and 6 used 75% of the dataset for training. In the experiments, the data points were randomly selected for each programming level. In experiment 7, 100% of the dataset was used for training and evaluation.

## 4. Results

The results, in Table 3, have shown from the experiments that the NFFPB model has a satisfactory improvement over the existing backfiring conversion ratios in MMRE and MMER. There was no conclusive evidence that neuro-fuzzy improves PRED because the results had positive, negative, and zero improvement.

The improvements were small in experiment 1 and 2 because of the size of the available dataset. Certain programming languages had a low number of training points which resulted in a small calibration. However, regardless of the setback, it was shown in experiments 5, 6, and 7 that there was a larger average MMRE and MMER improvement over the 50% random test results.

In experiments 4 and 7, the MMRE improvement was large. MMRE performs better when evaluated with large sized projects because MMRE favors underestimation. In experiment 7, there was a large improvement because the same data points were used for training and evaluation.

*Table 3 – Experiment Results*

| Experiment | Training Samples | Test Samples | MMRE Improvement | MMER Improvement | PRED(25%) Improvement | PRED(50%) Improvement |
|---|---|---|---|---|---|---|
| 1 | 121 | 120 | 4.29 | 7.89 | 0.00 | 0.00 |
| 2 | 121 | 120 | 1.02 | 11.35 | 2.56 | 0.00 |
| 3 | 121 | 120 | 10.77 | 7.30 | 0.00 | 7.41 |
| 4 | 121 | 120 | 15.00 | 6.00 | -2.86 | 1.61 |
| 5 | 180 | 61 | 7.35 | 9.26 | 0.00 | -3.23 |
| 6 | 180 | 61 | 8.45 | 5.43 | 10.53 | 3.33 |
| 7 | 241 | 241 | 26.39 | 9.29 | 0.00 | 1.71 |

There were negative improvements in PRED in experiments 4 and 5. The negative PRED

improvements indicate that during training some of the accurate project estimates were sacrificed in order to minimize the overall MMER error.

Figure 5 shows the inverse curve relationship between the calibrated weights from experiment 7 and the language level, which is similar to the curve in Figure 1. Similarly, the calibrated weights from other experiments have a similar curve.

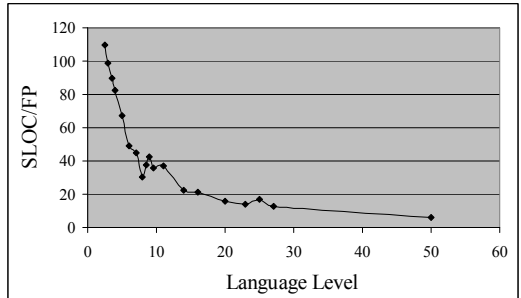

*Figure 5 – Calibrated conversion ratios versus language level*

## 5. Conclusion

The conclusions that were drawn from the NFFPB model were that it solved the issues of backfiring conversion and that it reduced the error of the size estimates. The issues with backfiring were that the conversion ratios had a large range and were generic.

Seven experiments were conducted to compare between NFFPB's calibrated ratios and SPR's conversion ratios. Overall, the NFFPB model outperformed SPR's conversion ratios in MMRE and MMER criteria. However, Neuro-fuzzy did not demonstrate any improvement in PRED.

The size of the project data available for evaluation weakened the conclusions being drawn because if more data became available, a larger improvement would show. The small training data hindered the neural network from minimizing errors and the improvements were small because of the biases in the MMRE and MMER criteria. In this study, the model tried to satisfy both criteria, which resulted in only obtaining local minimum error points.

## 6. Acknowledgment

Justin Wong would like to thank his supervisors and referees for their helpful comments. Furthermore, he would also like to thank SPR and ISBSG for providing research data.